\documentclass[lettersize,journal]{IEEEtran}
\usepackage{amsmath,amsfonts}
\usepackage{algorithmic}
\usepackage{algorithm}
\usepackage{array}
\usepackage[caption=false,font=normalsize,labelfont=sf,textfont=sf]{subfig}
\usepackage{textcomp}
\usepackage{stfloats}
\usepackage{url}
\usepackage{verbatim}
\usepackage{graphicx}
\usepackage{cite}
\usepackage{acronym}
\newacro{ADC}{Analog Digital Converter}
\newacro{AMR}{anisotropic magnetoresistance}
\newacro{EC}{Engine Controller}
\newacro{MCU}{Micro Controller Unit}
\newacro{TPS}{Throttle Position Sensor}
\newacro{TVA}{Throttle Valve Actuator}
\newacro{TbWS}{Throttle-by-Wire-System}
\begin{document}

\title{Low-Cost Throttle-By-Wire-System Architecture For Two-Wheeler Vehicles}

\author{ Jannis Kreß, Jens Rau, Hektor Hebert,  Fernando Perez-Peña, Karsten Schmidt, and Arturo Morgado-Estévez

\thanks{
The herein presented research was funded and supported by the University of Cadiz and Frankfurt University of Applied Sciences. The authors acknowledge the valuable support of Peugeot Motocycles Deutschland GmbH.\\ 
Jannis Kreß$^{1,2}$ (jannis.kress@fb2.fra-uas.de), Jens Rau$^{1}$, Hektor Hebert$^{1}$, Fernando Perez-Peña$^{2}$, Karsten Schmidt$^{1}$, Arturo Morgado-Estévez$^{2}$. Affiliations: $^{1}$ Department of Computing and Engineering, Frankfurt University of Applied Sciences, Germany. $^{2}$ Department of Automation, Electronics and Computing Architecture and Networks, University of Cadiz, Spain.}}%



\maketitle

\begin{abstract}
This paper investigates the performance of a low-cost \ac{TbWS} for two-wheeler applications. Mopeds/scooters are still restricted environmentally harmful. Throttle-by-Wire-Systems can contribute to environmental protection by replacing conventional restrictors. Its consisting of an \ac{AMR} throttle position sensor and a position controlled stepper motor driven throttle valve actuator. The decentralized throttle position sensor is operating contactless and acquires redundant data. Throttle valve actuation is realized through a position controlled stepper motor, sensing its position feedback by Hall effect. Using a PI-controller, the stepper motors position is precisely set. Both units are transceiving data by a CAN bus. Furthermore, fail-safe functions, plausibility checks, calibration algorithms and energy saving modes have been implemented. Both modules have been evaluated by Hardware-in-the-Loop testing in terms of reliability and measuring/positioning performance before the system was integrated into a \textit{Peugeot Kisbee 50 4T} (Euro 5/injected). Finally, the sensor unit comes with a measurement deviation of less than 0.16\% whereas the actuator unit can approach throttle valve positions with a deviation of less than 0.37\%. The actuators settling time does not exceed 0.13s while stable, step-loss free and noiseless operation. 
\end{abstract}


\section{Introduction}
\IEEEPARstart{I}{n} the automotive sector, \acp{TbWS} have been state of the art since the late 1980s \cite{BMW_TbW}. To enable assistance and safety systems, the intervention into longitudinal vehicle dynamics, mechanic connections are being replaced by electronic signals. A \ac{TPS} measures the throttle grip/pedal position and provides a digital value to the \ac{EC}. The throttle valve is controlled by an actuator using the control signal provided by the ECU \cite{TbW}. It took years until the benefits of \acp{TbWS} outweighed the technical challenges for the use in motorcycles. The need for additional systems with extra weight and space requirements was evident. With the arrival of advanced driver assistance systems, the advantages of \acp{TbWS} became obvious. In the past, \acp{TbWS} had been mainly applied to high performance sport bikes \cite{BMW_S1000RR_TbW}. Meanwhile, they are also being equipped to mid-class motorcycles, although their use on small motorcycles or scooters has not been established yet. This can be argued by additional costs with little added value. In this paper we present a \ac{TbWS} architecture that opens up the optimization of the motor operation point \cite{ETC_Peugeot} by using an innovative, alternative and more ecological restriction method. Modern combustion powered scooters must fulfill Euro 5 requirements \cite{EURO5}. Therefore, they are equipped with gasoline injection and a catalytic converter \cite{Peugeot, Kymco}. For effective exhaust gas after-treatment, the combustion air ratio ($\lambda$) must be close to 1 \cite{lambda}. By legal requirements, scooters (50 cc) must be restricted to 45 kph. This is achieved by shifting the ignition timing, since reducing the amount of fuel injected would cause $\lambda$ to become larger 1 (lean mixture). The low-cost \ac{TbWS} can contribute greatly to reduce fuel consumption and improving exhaust gas composition. Consequently, the potential of CO$_{2}$ savings and reduction of pollutants is significant. The system will be placed in the engine compartment to simplify vehicle integration and minimizing the electronic effort by using the original throttle cable. In this way, the \ac{TbWS} could also be retrofitted with little effort. The approach is novel within this vehicle class and the belonging powertrain type.

\subsection{Background on Throttle Position Sensors}
\begin{figure}[!h]
\centering
\includegraphics[width=8cm]{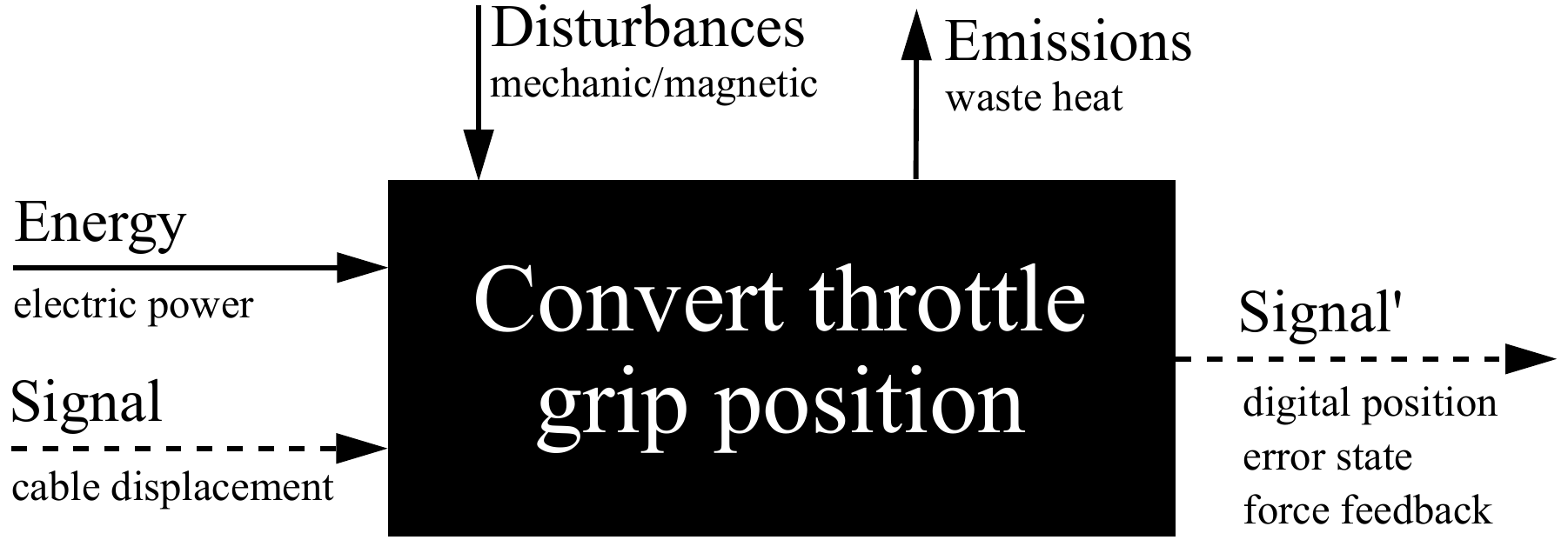}
\caption{\ac{TPS} blackbox schematic}
\label{blackbox}
\end{figure}

The \ac{TPS} transfers the physical input of the rider to an electronic signal while monitoring for plausibility. Figure \ref{blackbox} is giving a black box approach for \acp{TPS}, showing inputs and outputs. The \ac{TPS} needs power supply and converts the rotary/translatory physical input from the throttle grip. External influences like vibrations, electro-magnetic and magnetic fields must also be considered. The unit outputs a digital throttle position via vehicle bus for further processing.

Positions can generally be measured with relative and absolute sensors. Relative sensors are not considered due to their comparatively high error rate. With an absolute measurement, the sensor directly provides the angular position depending on its current location. The current state of the art is described by resistive, inductive and magneto-static sensor principles. Slip potentiometers are the cheapest and simplest method, but cause long-term errors due to their contact-based design \cite{Poti}. Inductive sensors measure contactless by angle-dependent induction changes in the coils. A redundant layout is difficult to design due to the larger and more complex setup. Since inductive sensors are based on alternating current effects, evaluation by a microcontroller is only possible with complex measurement circuitry \cite{InductiveSensors}. For measuring the strength or orientation of magnetic DC fields, magneto-static sensors are used. Most commonly applied are galvano-magnetic effects like Hall, Gaussian and magneto-resistive effects. AMR and GMR allow standstill detection and distinction of rotation direction \cite{Paper_magnSens}. In contrast to inductive sensors, they can be miniaturized better and manufactured at low cost. The Hall effect is the best known magneto-static effect but the sensors characteristic is related to temperature fluctuations and aging. To avoid saturation and under-run, the magnetic field strength must be within a defined range. AMR and GMR sensors measure absolute angular positions within a 180$^\circ$ (GMR: 360$^\circ$) range by a varying resistance related to the orientation of a magnetic field. Sensor elements are set up in a Wheatstone arrangement giving a sinusoidal output signal 
\cite{AMRsetup}. Placing two AMR-elements in 45$^\circ$ angular offset, a position can be calculated by applying the \textit{arctan} function. In contrast to the GMR sensor, the AMR effect can also handle strong magnetic fields while operating in saturation. Therefore, the AMR sensor can be operated independently of magnetic intensity fluctuations. Temperature related non-linearity of max. 3\% must be taken into account during measurement \cite{Paper_AMRsens}. Due to modern manufacturing processes, AMR sensors can be manufactured redundantly on tiny chips.


\subsection{Background on Throttle Valve Actuators}
\begin{figure}[!h]
\centering
\includegraphics[width=6.5cm]{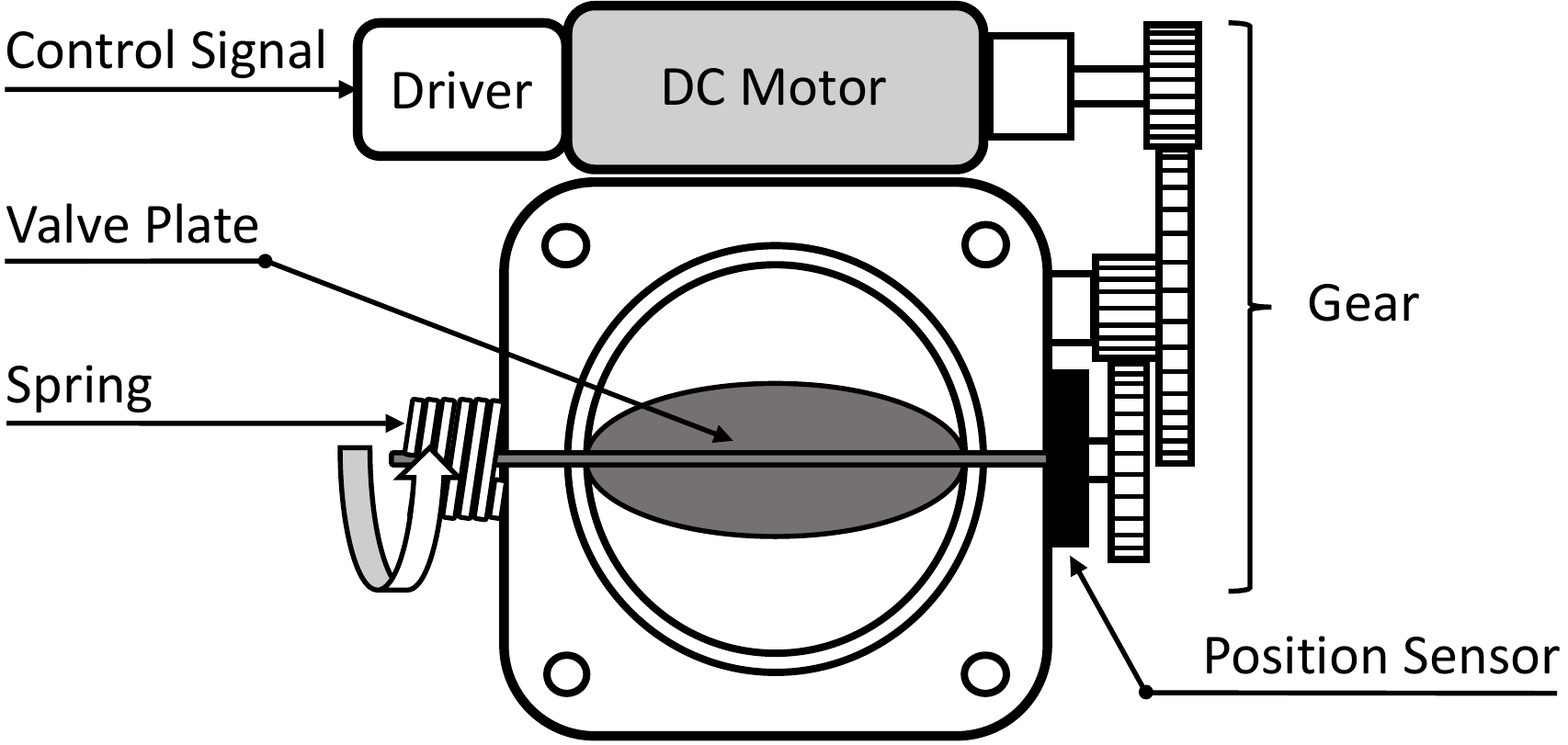}
\caption{\ac{TVA} schematic}
\label{TVA_scheme}
\end{figure}

Nowadays to manipulate the throttle valve with opening angles provided by the \ac{EC}, a \ac{TVA} is adapted to the valve shaft. Typically, a DC motor with a multi-stage transmission provides the required torque. For setting a desired position, a position control algorithm uses the feedback of an absolute angular position sensor. In the event of failure, a torsion spring moves the valve back to its initial state. Further, a motor driver is needed to interface the required set position from the \ac{EC} and to supply the motor with electrical energy. Figure \ref{TVA_scheme} is giving the basic schematic of a \ac{TVA}. 

Previous research was often based on a combination of DC motor, planetary or geared transmission, torsion spring and angular potentiometers for position measurement \cite{TVA_ETC_SportMotorbike, TVA_TbW_Ctr_SportMotorbike,TVA_ServoCtrDesign}. Beside the hardware setup, the research effort mostly is put on the position control algorithm. Even if \acp{TVA} appear trivial in terms of its design, mechanic and electrical non-linearity can lead to complicated control behavior \cite{Elec/Mech_NL}. Controlling \acp{TVA}, the upcoming throttle valve friction and in case of an integrated spring the belonging torque must be overcome. For that reason a common approach is a model-based control strategy, designed on the specific spring and friction characteristics. Setting the position is mostly realized through friction compensated PID-controllers, manipulating the motors supply voltage through a PWM signal \cite{TVA_ETC_SportMotorbike, TVA_TbW_Ctr_SportMotorbike, TVA_Nonl_PID}. Alternative approaches such as neural network based self-learning control \cite{NN_control} and adaptive sliding mode controllers have been applied successfully \cite{TVA_AdapCtr,TVA_AdapCtr_2,TVA_AdapCtr_3}. Further, linear parameter varying controllers \cite{TVA_Ctr4} or strategies such as minimal controller synthesis \cite{TVA_Ctr5} could also lead to high control performance. In context of automotive \acp{TVA}, a single, comparatively large throttle valve is attached independently from the number of cylinders. In motorcycles the number of throttle valves often corresponds to the number of cylinders. More complex mechanics are required and proportionally more friction effects can occur. In case of the \ac{TbWS} investigated, one throttle valve for single-cylinder engines is used.

\section{Development of \ac{TPS} \& \ac{TVA}}
Since a \ac{TbWS} is subdivided into the previously introduced components (\ac{TPS} \& \ac{TVA}), subsequent development steps are described accordingly. The overall requirements include the criteria of low component and manufacturing costs, based on the small vehicle class. Moreover, a system accuracy of at least 1\% of the throttle travel with a max. settling time of 200ms is assumed. Requirements are shown in table \ref{Tab_SysReq}. Beside dynamic performance, fail-safe requirements are fundamental in automotive industries such as a redundant and contactless \ac{TPS}. On top, failure detection of both components and the resulting handling of error conditions need to be managed.   

\begin{table}[h]
\begin{center}
\caption{System requirements}
\label{Tab_SysReq}
\begin{tabular}{l c c c }
\hline             & \ac{TPS}       & \ac{TVA}  & \textbf{\ac{TbWS}}         \\ \hline
Bus rate (Hz)      & 50             & 50        & \textbf{50}                \\ 
Accuracy (\%)      & 0.5            & 0.5       & \textbf{1.0}               \\ 
Settling time (ms) & 50             & 150       & \textbf{200}               \\ 
Displacement (deg) & 180            & 69        & \textbf{180/69 (I/O)}      \\ \hline
\end{tabular}
\end{center}
\end{table}
\vspace{-0.5cm}

\subsection{Throttle Position Sensor}
Basically, the sub tasks can be grouped into mechanics, electronics and software. Electronic aspects are the provision of electrical power, redundant detection of the throttle grip position and the consecutive pre-amplification. Then, both measured values are digitized. Before conversion, the now digital values need to be normalized with the previously measured calibration values by software. The results are checked for plausibility and sent to the CAN bus. From mechanical point of view, foreign matter must be retained by a housing and components must be shielded. In addition, the throttle cable must be guided and limited through end stops, while a counterforce must be built up as driver feedback.

\begin{figure*}[!b]
\centering
\includegraphics[width=14.5cm]{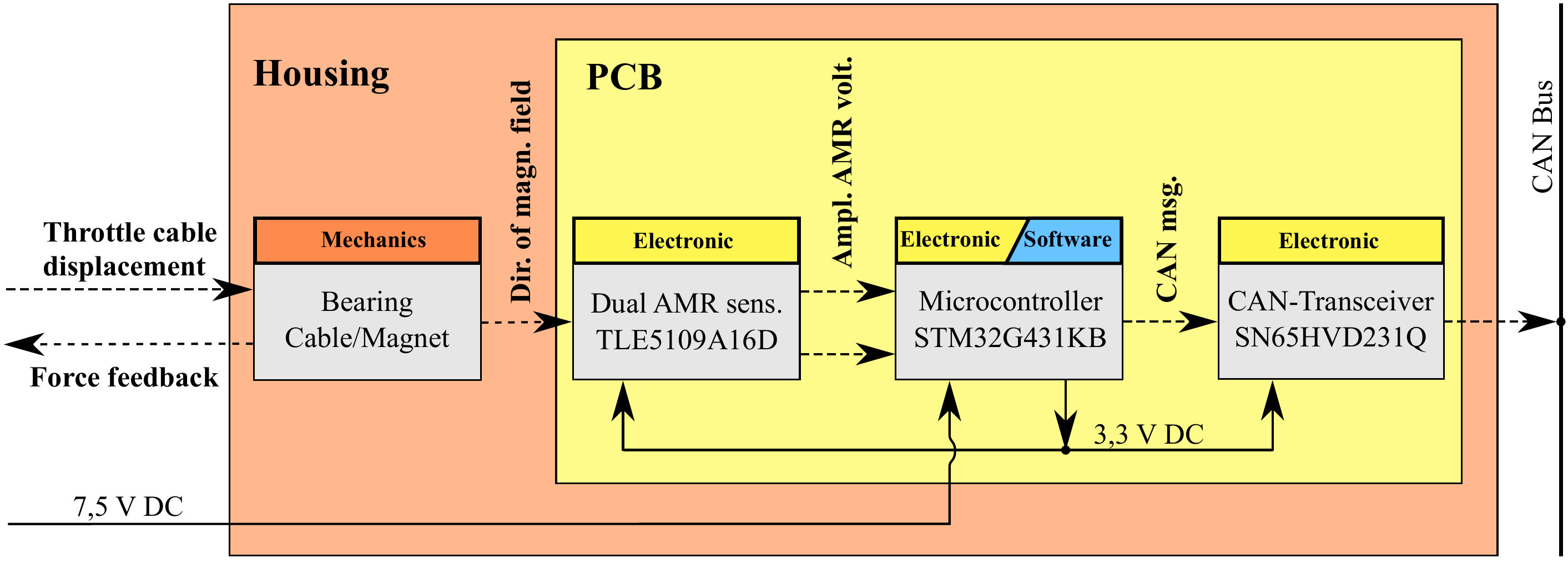}
\caption{\ac{TPS} architecture}
\label{TPS_Architecture}
\end{figure*}

\textbf{System Architecture:}
In order to implement the above-mentioned sub-functions, the following architecture results for the \ac{TPS}. Figure \ref{TPS_Architecture} shows the merging into one assembly of housing and PCB, covering the introduced levels. After considering the sensor characteristics, the decision was made of an cost-effectiv AMR sensor IC (\textit{TLE5109A16D E1210}) from \textit{Infineon Technologies AG} \cite{AMR_datasheet}. Placing the \ac{TPS} directly within the engine compartment, represents a harsh operating environment. This requires a robust sensor which is not susceptible to electromagnetic interference fields and operates properly even under highly dynamic loads. The original 69$^\circ$ rotation angle can be optimally changed to the 180$^\circ$ by adjusting the cable pulley. Due to position measurement under magnetic saturation, strong magnetic fields can be applied minimizing the susceptibility to interference. By its design the measurement principle is contactless and the IC is equipped with redundant electrically isolated AMR measuring bridges. The measurement signals are internally amplified and output as analog voltages. Here, the output signal can optionally be tapped differentially. These are being processed by a compact \textit{STM32-G431KB} \ac{MCU} which features two symmetrical digital conversions, one differentially. To perform trigonometric calculation more efficient, the processor has the option of hardware-acceleration by the  CORDIC. Scaling, conversion and plausibility checks are done by software before CAN message creation. Finally, a CAN transceivcer is executing the CAN communication including the from the \ac{MCU} provided \ac{TPS} message.\\

\begin{figure}[!h]
\centering
\includegraphics[width=8.8cm]{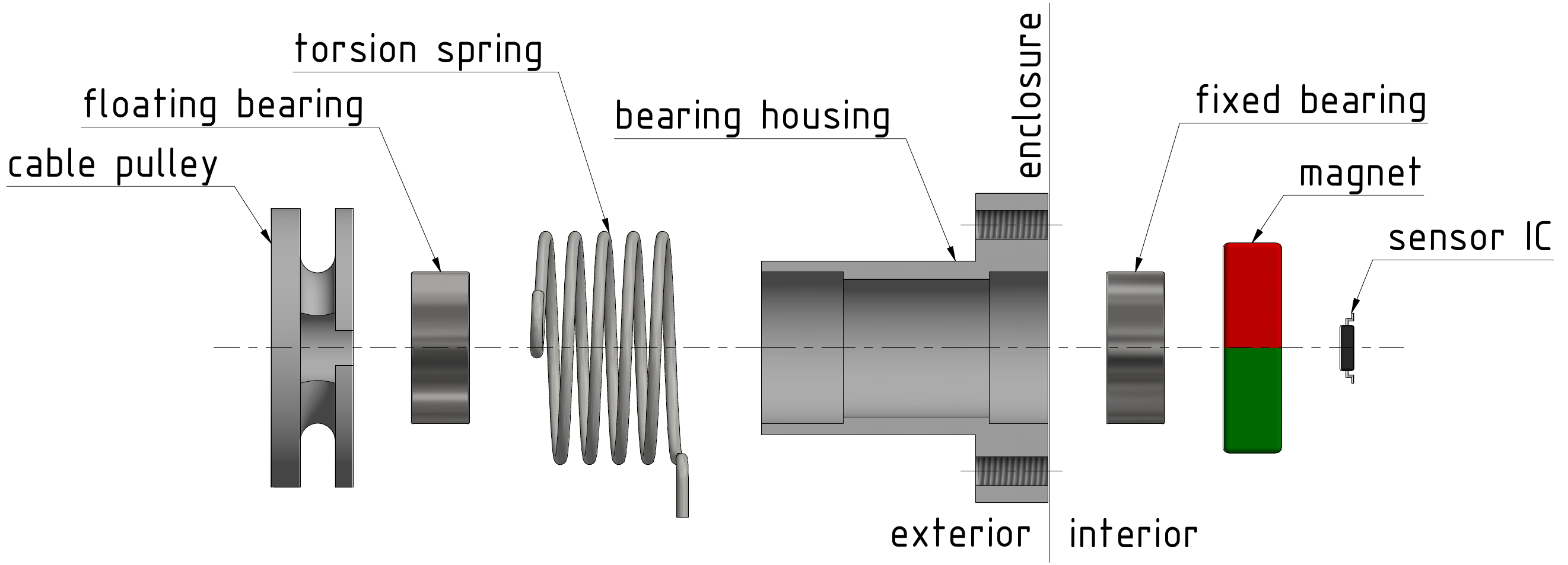}
\caption{\ac{TPS} mechanics}
\label{TPS_Mechanics}
\end{figure}

\textbf{Mechanical Implementation:}
To measure the angular position of an axis, a permanent magnet needs to be placed centered above the AMR-IC at the end of shaft. The throttle cable is picked up by a pulley. Its circumference has to be designed in such a way that the cable travel makes optimum use of the sensor measuring range. Manufacturing tolerances (3\%), positional inaccuracies within the chip/on the board (5\%), and error bands for error detection (5\%) were considered \cite{AMR_datasheet}. Consequently, 148$^\circ$ can be effectively used of the measurement range (180$^\circ$). Good guidance of the shaft is achieved with a floating/fixed bearing arrangement. In order to generate the required throttle feedback, the bearing features a torsion spring, which applies the torque for a restoring force. Figure \ref{TPS_Mechanics} illustrates the shaft arrangement. Besides handling the mechanical input, a housing is to be designed. External environmental influences must be prevented from the \ac{TPS} while the PCB and the bearing unit must be integrated.

\textbf{Electronic Implementation:}
The PCB ensures a secure electrical connection of all electronic components, enabling space-saving arrangement. The assembled PCB is mounted in the sensor housing and represents the electronic system level. Amplified sine/cosine signals are provided by the AMR-IC in redundant way. One measuring bridge is to be read out differentially in order to minimize noise due to electromagnetic interference. In addition, due to better EMC low-pass filtering, support capacitors are connected to the signal outputs. The \ac{MCU} provides sufficient \ac{ADC} channels for differential conversion of a single measuring bridge. By non-differential conversion of the second one, a reference for plausibility check is generated.\\

\textbf{Software Implementation:}
For secure and immediate engine control, the throttle position is to be detected with a sampling rate of min. 50 Hz. Two \acp{ADC} digitize the measured values, which are directly transferred to the memory by the Direct Memory Access (DMA) controller. Thereby, the max. angular velocity of the pulley must be taken into account, which has been determined experimentally to $\omega_{\text{max}}=9\frac{\circ}{\text{ ms}}$. In order to meet the accuracy requirement of 1\%, the occurring differences in the conversion time, resulting from the sequential conversion, must be respected. The longest possible conversion time is represented in (\ref{eq.1}).
\begin{equation}
     \textit{t}_{\text{\textit{conv,max}}}=\frac{\textit{Angle} _{max}}{100\cdot\omega _{max}}
    =\frac{148^\circ}{100 \cdot9\frac{\circ}{\textit{ms}}}=164\mu\textit{s} 
    \label{eq.1}
\end{equation}
Transferring $\textit{t}_{\text{\textit{conv,max}}}$ into CPU cycles, equation (\ref{eq.2}) is used.
\begin{equation}
    \textit{t}_{\text{\textit{conv}}(52\text{\textit{MHz}),\textit{max}}}=164\mu\text{\textit{s}}\cdot52\frac{\text{\textit{cycles}}}{\mu\text{\textit{s}}}=8528\text{\textit{cycles}}
    \label{eq.2}
\end{equation}
It is necessary to balance sampling time and over-sampling (multiple measurement incl. averaging by hardware). A sampling time of 640.5c with eight-fold oversampling was determined experimentally. The precise \ac{ADC} conversion time can be determined with (\ref{eq.3}) \cite{STM_datasheet}. 
\begin{equation}
    \begin{split}
     \textit{t}_{\textit{conv}}=\textit{t}_{\textit{s}}[\textit{cycles}]+\textit{resolution}[\textit{bits}]+0.5 \\
     \textit{t}_{\textit{conv}}= (640.5+12+0.5) \cdot8 = 5224 \textit{cycles}  
    \end{split}
    \label{eq.3}
\end{equation}
The cycles required are less than for $\textit{t}_{\text{\textit{conv}}(52\text{\textit{MHz}),\textit{max}}}$ and the precision requirements can be fulfilled. To be energetically efficient, the \acp{ADC} are triggered by timer based interrupts and storage is processor-independent via the \acp{MCU} DMA. After \ac{ADC} conversion, calibration calculations are applied to convert the measurements to normalized values within range. The AMR output signals can be described by (\ref{eq.4A}) and (\ref{eq.4B}).
\begin{subequations}\label{eq:4}
\begin{align}
\textit{SIN}_{\textit{AMR}}= \textit{Y} = \textit{A}_{\textit{Y}}\cdot\textit{sin}(2\cdot\alpha)+\textit{O}_{\textit{Y}}  \label{eq.4A}\\
 \textit{COS}_{\textit{AMR}}= \textit{X} = \textit{A}_{\textit{X}}\cdot\textit{sin}(2\cdot\alpha)+\textit{O}_{\textit{X}} \label{eq.4B}
\end{align}
\end{subequations}
The function \textit{atan2} calculates the angle of rotation $\alpha$ from the ratio of both output signals. They must be transformed to the same value range of [-1,1] by absolute displacement and scaling of the measured values given in (\ref{eq.5}) (likewise for \textit{X}).
\begin{equation}
    \textit{Y}_{1}=\textit{Y}-\textit{O}_{\textit{Y}}\ \ ;\ \ \textit{Y}_{2}=\frac{\textit{Y}_{1}}{\textit{A}_{\textit{Y}}}
    \label{eq.5}
\end{equation}
Now, angle of rotation can be calculated from both normalized values: $\alpha = atan2(X2, Y2)$. To solve the trigonometric function, a Taylor series is used to approximate the result in a computationally intensive way. This is remedied by the \acp{MCU} CORDIC co-processor, which approximates trigonometric functions at hardware level in less processor cycles. To avoid amplitude shifts and offsets, an end-of-line calibration can be performed, which requires the complete measuring range to be scanned. Offset (\textit{O}) and amplitude maxima (\textit{A}) can be calculated from the recorded values by means of (\ref{eq.6}).  
\begin{equation}
    \textit{A} = \frac{\textit{Max}-\textit{Min}}{2}\ \ \ ;\ \ \ 
    \textit{O} = \frac{\textit{Max}+\textit{Min}}{2}
    \label{eq.6}
\end{equation}
Before the throttle position is transmitted, the redundantly acquired measured values must be checked for plausibility. The range of values and the signals magnitude can be verified as properties of the individual signals. Redundantly acquired measurements are checked by signal or angle comparison. With equations (\ref{eq.7A})/(\ref{eq.7B}) the by the manufacturer specified value range of the AMR sensors can be verified with regard to its minimum/maximum. 
\begin{subequations}\label{eq:7}
\begin{align}
\begin{split}
\textit{Min}_{\textit{diff}} = \textit{O}_{\textit{diff,min}}-\textit{A}_{\textit{diff,max}}-\textit{V}_{\textit{Noise}} \ \ \ \ \\
\textit{Min}_{\textit{diff}}= -200 \textit{mV}-2.6\textit{V}-5\textit{mV}=-2.805\textit{V}  \label{eq.7A}
\end{split}\\
\begin{split}
 \textit{Max}_{\textit{diff}}= \textit{O}_{\textit{diff,max}}+\textit{A}_{\textit{diff,max}}+\textit{V}_{\textit{Noise}}\\
\textit{Max}_{\textit{diff}} = 200 \textit{mV}+2.6\textit{V}+5\textit{mV}=2.805\textit{V} \label{eq.7B}
\end{split}
\end{align}
\end{subequations}
To detect erroneous measured values, the magnitude check can also be beneficial. For this purpose, the phase shift of a quarter period (orthogonality) is exploited by plotting the sine over the cosine function. The resulting location vector traverses a unit circle with the radius (\textit{r}) for changes of the rotation angle ($\varphi$). Based on the max. permissible angular error (1.48$^\circ$) an upper limit of $\Delta$r = 2.5\% results. Finally, the values are compared with the reference of the second AMR sensor with regard to the concrete value and the resulting angular offset. Again, the angular error must not exceed 1.48$^\circ$ and the deviation must be less than $\Delta SIN_\textit{max} = 0.26$ (tolerance 5\%). These analyses allow to detect short circuits, cable breaks, defects of the measuring bridges and electromagnetic interference. After plausibility checks, a CAN message can be created featuring the digitized throttle position, message counter and error status of the sensor unit. For a precise position transmission a 10 bit resolution has been chosen, resulting in a percentage quantity with one decimal place precision. 

\subsection{Throttle Valve Actuator}
Developing the \ac{TVA} also requires decomposition into electronics, software and hardware level. Electronic aspects are the conversion and amplification of the stepper motor control signal, the contactless measurement of the valve position and the belonging preamplification/digitization. The measured position needs to be checked for plausibility by software and will serve as control feedback. A position control must set the requested throttle valve position, received by the bus. For fail-safe reasons the actuators operation needs to be monitored regarding stability, precision and failure. To guarantee smooth/energy-efficient operation and good engine idle speed behavior, an online calibration and energy saving mode has to be implemented. From the mechanic point of view, foreign matter must be retained by a housing and components must be protected. The powertrain must be designed consisting of a motor, a proper transmission and the position sensor adaption, taking into account error-security. \\    

\begin{figure*}[!b]
\centering
\includegraphics[width=13cm]{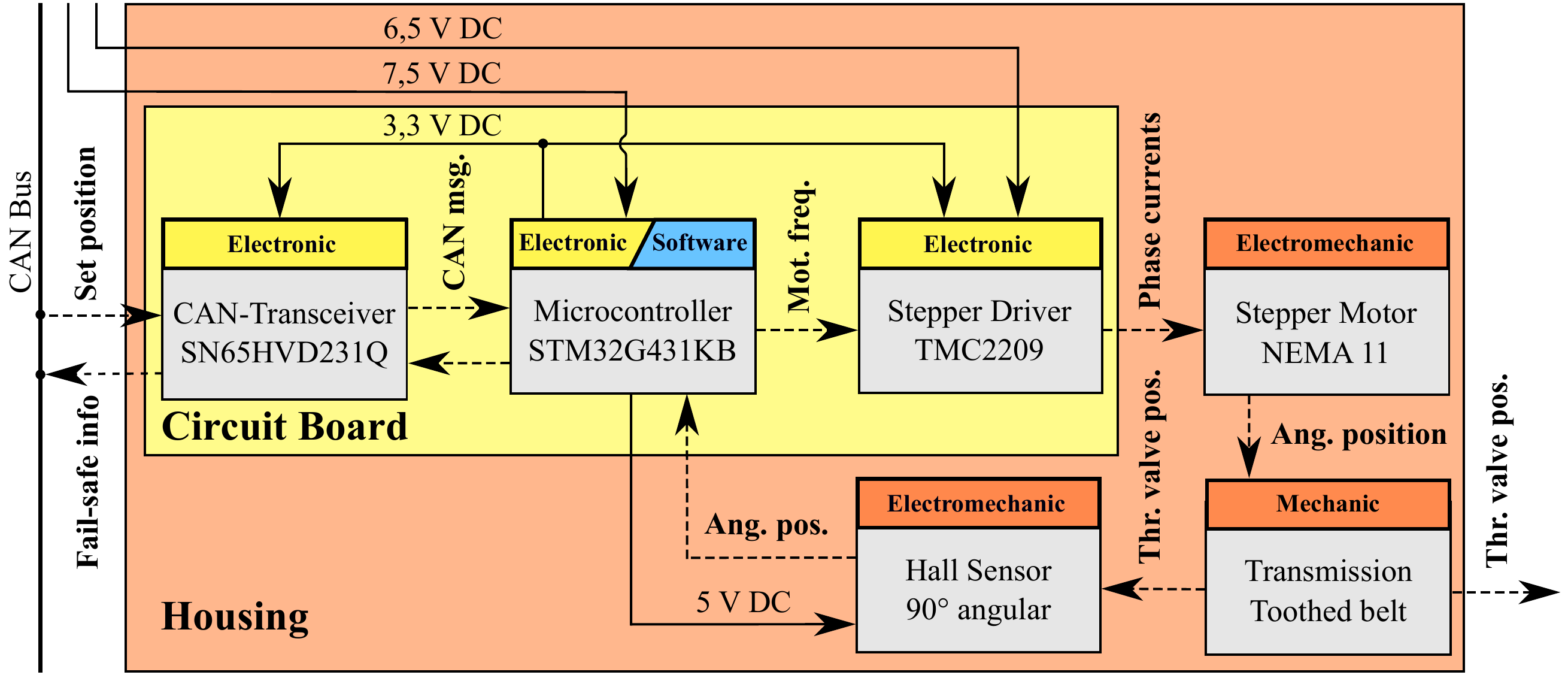}
\caption{\ac{TVA} architecture}
\label{TVA_Architecture}
\end{figure*}

\textbf{System Architecture:} 
Based on the given sub-functions, the following architecture can be derived, consisting of a circuit board and a housing like shown in figure \ref{TVA_Architecture}. The \ac{TVA} is supplied with two separately fused DC voltages to guarantee operation even in case of a collapsed \ac{TVA} powertrain. Output is the controlled throttle valve position. The electric motor selected is a \textit{NEMA 11} (\textit{HS20-0674S}) two-phase bipolar stepper motor \cite{NEMA11_datasheet}. In contrast to DC motor application, the use of a stepper motor means that friction effects can be neglected, as long as the torque provided is sufficient at the max. speed required. Thus, the need for a model-based controller is obsolete. Closed-loop position control allows readjustment in the event of step loss. To control stepper motors, the integration of a driver is mandatory. Therefore, a \textit{TMC2209} from \textit{TRINAMIC} had been utilized, offering high phase current supply and enabling precise motor control through micro-stepping \cite{TMC2209_datasheet}. In comparison to most \acp{TVA}, the one developed is coming without any torsion spring for fail-safe reasons. Typically, the spring is resetting the \ac{TVA} to idle position and keeps the engine running to drive all safety critical auxiliary units. In case of scooters/small motorcycles there are no such systems. All safety critical systems are driven manually, why fail-safe states can be realized by interrupting the ignition and simplifying the \acp{TVA} design that way. The torque provided by the motor is translated by a toothed belt transmission. Both output shaft ends are used to adapt the position sensor as well as the throttle valve. For position measurement, the decision was made for a non-contacting Hall effect single-turn position sensor (\textit{6127V1A90L.25FS}) from \textit{TT Electronics} \cite{HallSensor_datasheet}. With a min. resolution of $0.022^\circ$ and a max. linearity deviation of $\pm0.25\%$, the sensor meets the accuracy requirements of $\pm1\%$. Coming with a range of travel over $90^\circ$, this results in a max. measurement deviation of $\pm0.247^\circ$ (1\% = $0.7^\circ$).  It far exceeds the service life requirement of 500.000 (scooter service life) with 10 million cycles. Sensor read out, signal processing, motor and throttle valve position control are implemented on a similar \textit{STM32-G431KB}. Fail-safe functions, calibration and CAN communication are handled by the \ac{MCU} and a CAN transceiver.\\

\begin{figure}[!h]
\centering
\includegraphics[width=5cm]{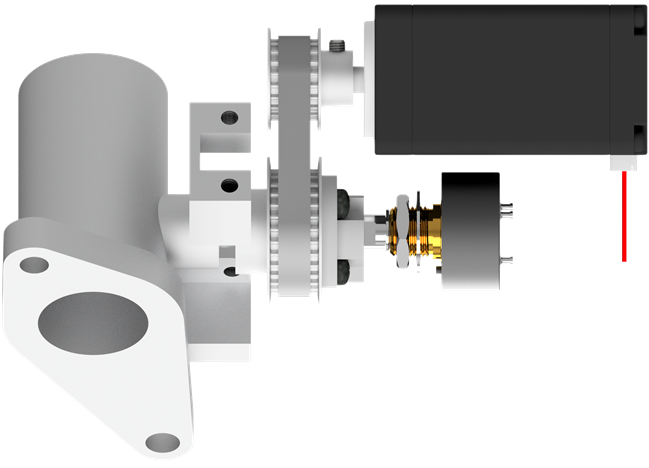}
\caption{\ac{TVA} mechanics}
\label{TVA_Mechanics}
\end{figure}

\textbf{Mechanical Implementation:}
In order to reach responsive system reactions, the max. original valve angle of 69$^\circ$ needs to be traveled in less than 200ms to prevent lagging throttle behavior. This results in a max. speed of 345$^{\circ}$/s. Caused by the high inertial characteristics of fan-cooled small-volume engines, the performance required is significantly lower than that of sports motorcycles. Stepper motors have torque losses with increasing speed, which is why the torque required for the highest operating point is a major criterion. To determine the characteristics with the chosen driver, a load series test was performed, showing a stable max. torque of 12 Ncm up to a speed of 360$^{\circ}$/s. The requirement of 5 Ncm was derived from measurements carried out by \textit{Scion-Sprays Ltd} and \textit{Peugeot Scooter, France} \cite{ETC_Peugeot}. Beside providing the needed performance, the torque must be transferred to the throttle valve shaft while enabling the positioning of the Hall sensor. For fail-safe reasons, the sensor should be mounted directly on the shaft to avoid measuring random position values in the event of a gearbox or motor failure. In case of a broken belt, the \ac{TVA} is still capable to distinguish between faulty or proper operation. Therefore, an adaption plate was designed, connecting the shaft with the sensor by the tooth belt pulley. With regard to the transmission, the decision was made of a slip-less toothed belt due to the low gear backlash and low maintenance. A 1.5-fold transmission ratio was selected, improving position accuracy. The resulting slight drop in max. torque can be accepted thanks to the motors torque surplus. Figure \ref{TVA_Mechanics} shows the mechanical \ac{TVA} setup.

\textbf{Electronic Implementation:}
The circuit board, shown in figure \ref{TVA_Architecture}, ensures a secure electrical connection of the CAN transceiver, \ac{MCU} and stepper driver. 6.5V and 7.5V supply voltages are provided by a central power supply, whereby the 6.5V line is individually fused for driving the motor. 3.3V supply for the CAN transceiver as well as the stepper drivers processing electronics are provided stabilized by the \ac{MCU} itself. The assembled circuit board is mounted at the throttle valve body and represents the electronic system level. An analog position signal is delivered by the externally placed Hall sensor, which is to be digitized by the \acp{MCU} \ac{ADC}. A 256-fold oversampling was figured out for best noise reduction. Stepper motors characteristics depend on the driver used. The chosen driver realizes an extremely smooth and yet powerful motor run through automatic micro-stepping (up to 256). Thus, the step-like rotation becomes harmonic and slow position changes can be realized noiseless.\\

\begin{figure}[!h]
\centering
\includegraphics[width=8.1cm]{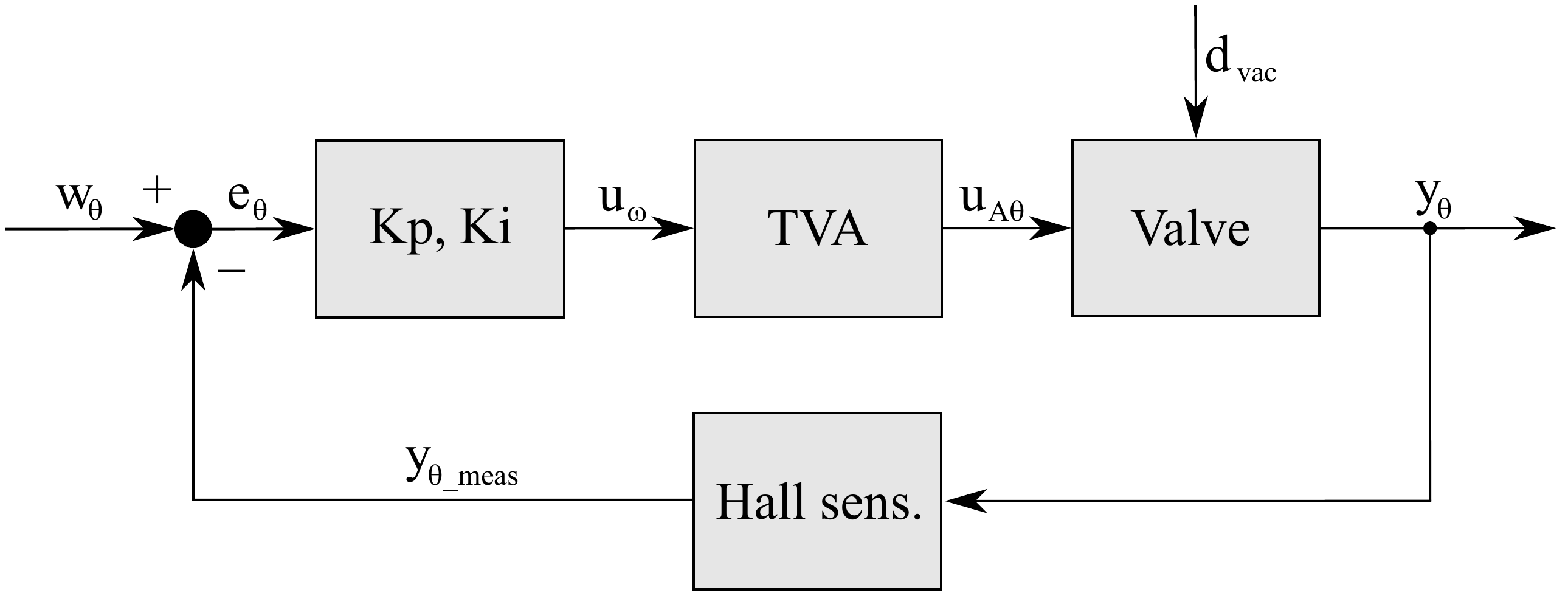}
\caption{\ac{TVA} control loop}
\label{TVA_ControlLoop}
\end{figure}

\textbf{Software Implementation:}
The \ac{TPS} request needs to be received and processed every 20ms (50Hz). For creating the final throttle valve set position, the received 10 bit quantity is transferred to an opening relation with one decimal place precision (0.0\% $\hat{=}$ idle state). At the same rate, the current throttle position, message counter and error status are sent to the bus for diagnostic purposes. Beside knowing the set position, its most important to measure the current valve position to implement a closed-loop control algorithm. Therefore, the digitized and averaged position value is also being transferred into a relative quantity (\%) with a precision of one decimal place. Afterwords, the position is again smoothed by applying a moving average filter to compensate lowest powertrain vibrations. Now all variables required for position control are known and the 1kHz clocked PI-control algorithm can be set up like shown in figure \ref{TVA_ControlLoop}. If stepper motors are operated without micro-stepping, their angular increments of 1.8$^\circ$ are the best possible control resolution and a P-controller is sufficient. The required position would always be achieved position controlled, assuming that the motor was sufficiently dimensioned. Step loss-free operation is the prerequisite. Here, the up to 256-fold micro-stepping results in a harmonic almost step-free motor rotation. To compensate small, longer lasting deviations, an I-controller was implemented. 

The required valve position (w$_\theta$) serves as input variable. After building the control deviation (e$_\theta$) by summing w$_\theta$ and the inverted Hall sensor feedback (y$_\theta$\textsubscript{\_meas}), the PI-controller outputs the manipulated variable u$_\omega$ in form of an angular velocity/step frequency.  Next, the \ac{TVA} adjusts the throttle valve by the actuator manipulated variable u$_{A\theta}$, which can be disturbed by the engine speed-dependent intake pressure (d$_{vac}$). Nevertheless, d$_{vac}$ is too weak to falsify the set position, since the motor torque is sufficient. The resulting valve position y$_\theta$ is measured by the Hall sensor and fed back. Beside the step frequency, the drivers enabling/disabling and direction of rotation must be preset by the \ac{MCU}.\\

In terms of fault tolerance, the \ac{TVA} must clearly be classified as safety-critical. In in the event of malfunction or defect, the longitudinal dynamics of the vehicle can be strongly influenced. Although the performance of scooters is still low that manually over-braking is possible, this is not to be expected. To detect failures like a broken transmission belt (1), stepper motor (2), motor driver (3), position sensor (4) or the \ac{MCU} (5), a fail-safe algorithm is to be developed. In case (1), (2) and (3) the \ac{TPS} set position will clearly differ from the Hall sensor acquired position. Therefore, the error is determined with two moving average filters (w$_\theta$ \& y$_\theta$\textsubscript{\_meas}). If the error exceeds a predefined threshold of 5\%, the error state is declared by the overflow of a fail-safe timer. In the event of a faulty/defect sensor (4), the predefined angular range can be monitored to detect mechanical issues like overshoots/undershoots or even a total failure by comparing the signal line to the supply voltage/GND. Even if one \ac{MCU} fails (5), the system must be able to induce a safe state. If the \ac{TPS} \ac{MCU} fails, the error will be detected by the \ac{TVA} due to the faulty message counter. Then, the actuator refuses commands and moves back to idle position. In the event of self-detected \ac{TVA} errors, the TPS is addressed by sending an error report, interrupting the ignition by means of a safety contactor. In case of total failure of the \ac{TVA} \ac{MCU} (5) the same effect occurs by detecting an outdated message counter.\\  

Stepper motors can consume the phase short-circuit current in energized standstill. For this reason, an energy-saving mode has been implemented that tolerates the smallest deviations while taking the control deviation into account. If the deviation is less than 0.15\% within a predefined time period, the motor is no longer energized by disabling the driver. That way, the current consumption and the generated waste heat can be significantly reduced, enabling a passive cooling strategy. In addition, a small permanent deviation of the set \ac{TVA} position can occur due to minimal temperature-related drifts of the Hall sensor. These minimal deviations do not have any remarkable effect on partial or full-load behavior, but the engines idle speed is already affected by 0.5\% deviations. To avoid this, a calibration function has been implemented which calibrates the actuator within 0.5s when the system is switched on. This process is repeated in larger time intervals. As soon as the vehicle does not receive a throttle command, the re-calibration is executed. For this purpose, a three-point controller was specially tuned, which determines the correct initial position depending on the measured unstable position when the physical lower throttle valve limitation is reached.

\section{Results}
The \ac{TPS} and \ac{TVA} are evaluated individually before both are tested in context of a \ac{TbWS}, including CAN bus communication. Both developed components are described and their performance is evaluated by means of dynamic/static precision and response time by performing dynamic tests. Finally, influences of the engine suction pressure are investigated.

\subsection{Evaluation of the \ac{TPS}}
\begin{figure}[!h]
\centering
\includegraphics[width=6cm]{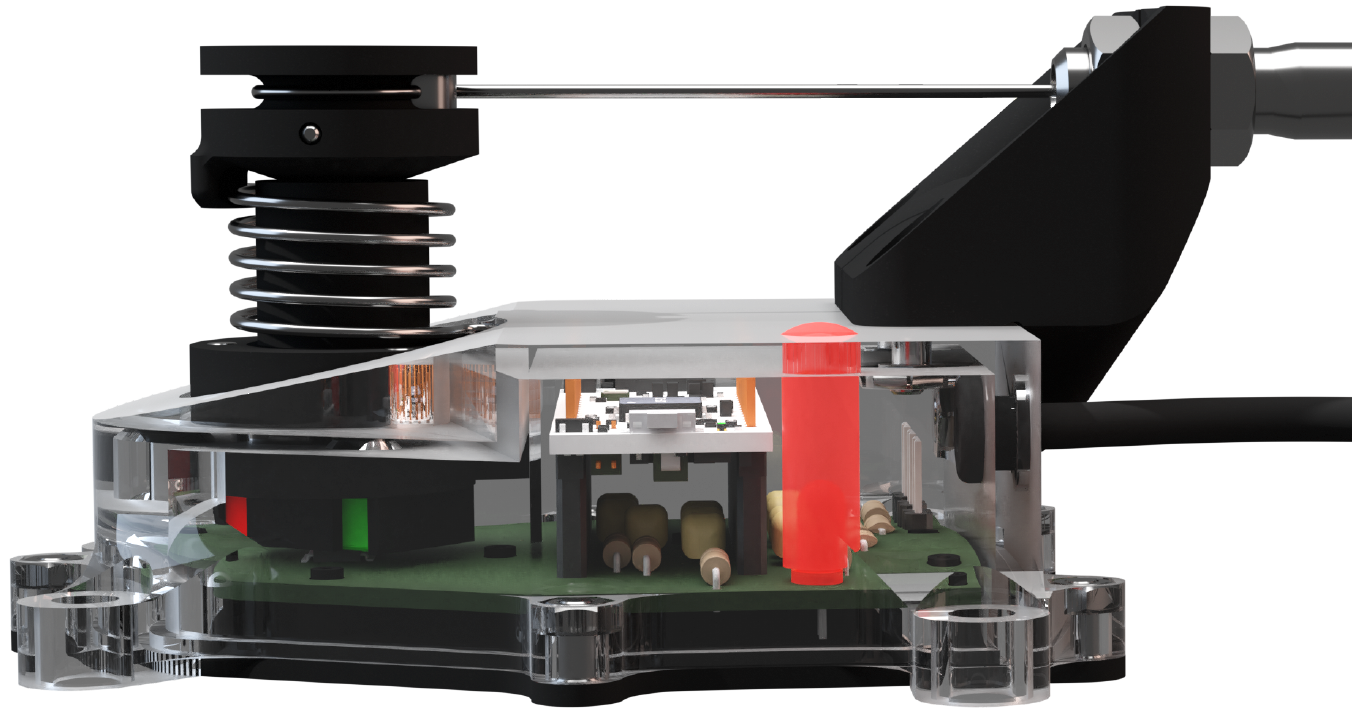}
\caption{\ac{TPS} assembly}
\label{TPSunit}
\end{figure}

A PCB was designed to merge all \ac{TPS} electronic components. Differential signal lines were guided in parallel with same length, to minimize the effect of electro-magnetic interference. To prevent deflection of the magnetic field, passive components such as capacitors and resistors were placed as far away from the sensor-IC as possible. A LED on the circuit board is used to indicate detected errors. Figure \ref{TPSunit} shows the \ac{TPS} assembly of the PCB, the housing and the bearing unit including the cable pulley pickup. By using a torsion spring, the original feedback force is encountered to the throttle cable. The unit is connected to a four-pole connector, providing the 7.5V supply voltage and the CAN bus access. The housing is closed and a circumferential seal protects the installed electronics from environmental influences.

\begin{figure}[!h]
\centering
\includegraphics[width=8.6cm]{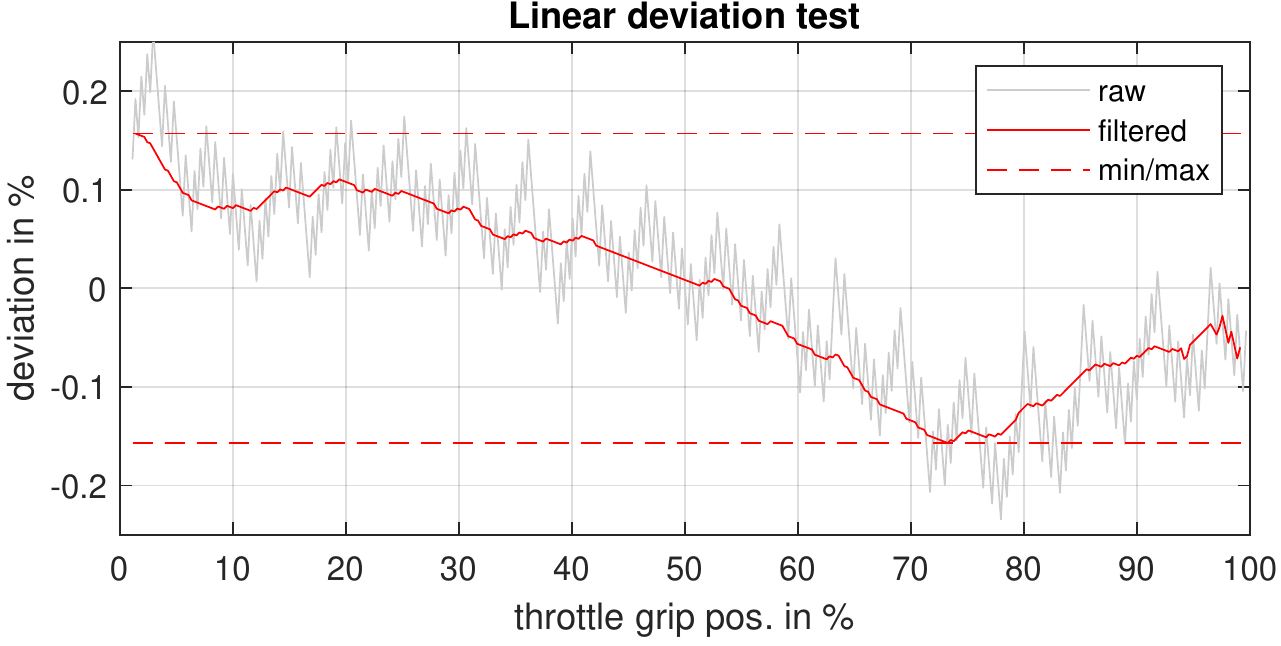}
\caption{\ac{TPS} open-loop deviation testing}
\label{TPS_deviation}
\end{figure}

For evaluating the \ac{TPS}, a HiL-test environment was set up where the position is varied by use of a stepper motor. Here, the digital throttle position was continuously set to randomly selected values (90 setpoints/min) by position control. The endurance test bench ran for a total of 60 h, where about 324.000 positions were set and 10.800.000 value pairs were processed. No errors occurred during the entire endurance test. 
The setpoint is adjusted by the motor with an accuracy of less then 0.1\%.
By means of the endurance test bench, no statement can be made about the measuring accuracy, why an open-loop test scenario was developed for determining the linearity deviation. A stepper motor rotates the sensor shaft step-wise, while position measurements are output. The angular position of the sensor was analyzed with a step width of approx. 0.38$^\circ$ (360$^\circ$/950 steps) for the entire measuring range. Figure \ref{TPS_deviation} shows the measurement deviation related to the reference position of the stepper motor. Since the deviation varies periodically, caused by tolerances in the steppers gearbox, this was corrected by applying a moving average filter. A deviation of $\Delta$dev$_{max} = 0.155\%$ results.

\vspace{-0.4cm}

\subsection{Evaluation of \ac{TVA}}
\begin{figure}[!h]
\centering
\includegraphics[width=8.6cm]{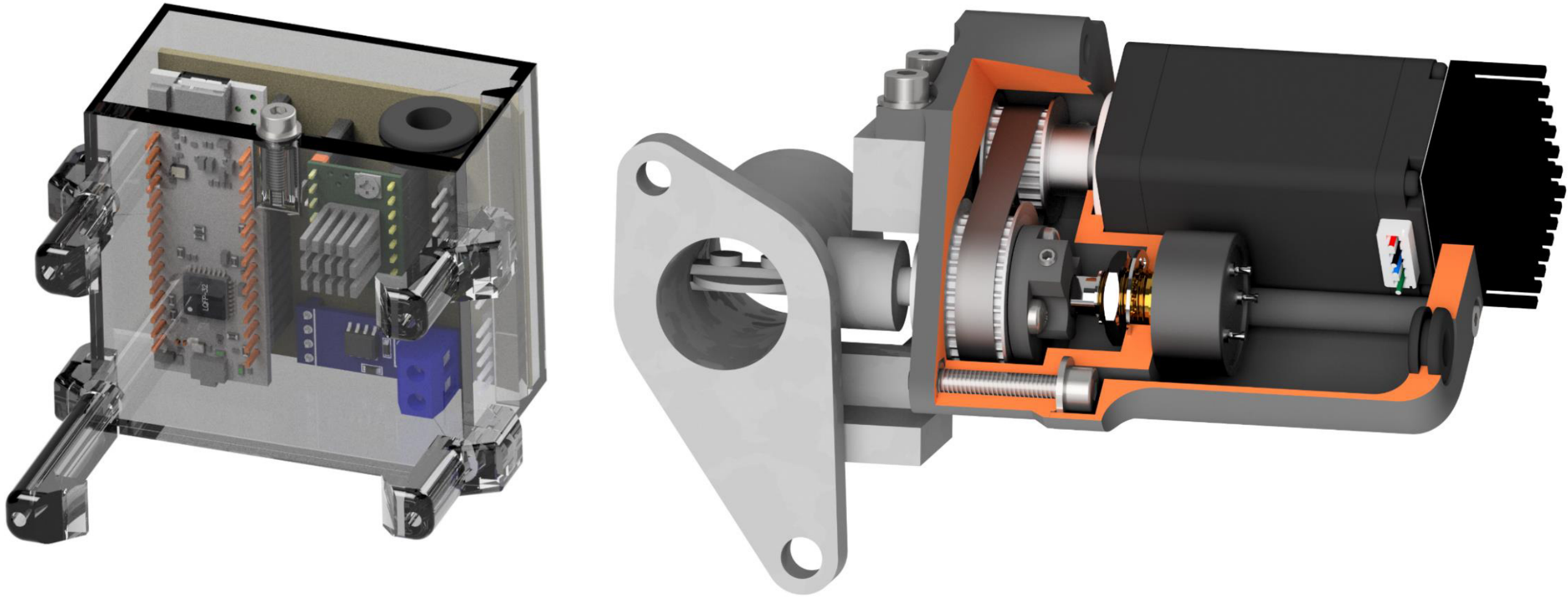}
\caption{\ac{TVA} electronic and mechanic unit}
\label{TVA_unit}
\end{figure}

Scooters do not come with a conventional rear swing arm, since the whole powertrain (engine, gearbox, wheel) is mounted movable. If the rear suspension compresses strongly, a large unit mounted on the throttle body would be destroyed. To prevent collisions with the scooters frame, the \ac{TVA} has been divided into smaller units. An electronic unit was designed, consisting of a circuit board carrying the \ac{MCU}, stepper driver and CAN transceiver which is integrated into a housing that fits on the backside of the original \ac{EC}. The unit is connected through a five-pole connector providing 6.5V supply for the stepper drivetrain, 7.5V supply for the \ac{MCU} and CAN bus access. A circumferential seal protects the electronics from environmental influences. The second unit is carrying the electromechanic and mechanic components like Hall sensor, stepper motor, transmission and adaption plates. Therefore, a three-stage design was developed, whose first stage realizes the adaptation to the throttle valve and the protection of the gearbox. The second is given by a base plate on which stepper motor, Hall sensor and toothed belt gear are installed. Third component is the housing cover, which enables passive cooling and cable feed-through. Splash water protection is sufficient and no seals are needed due to the robust components. Figure \ref{TVA_unit} shows the electronic (left) and mechanic unit (right). Both are connected with a seven-core cable, enabling sensor supply/readout and motor driving.

\begin{figure}[H]
\centering
\includegraphics[width=8.6cm]{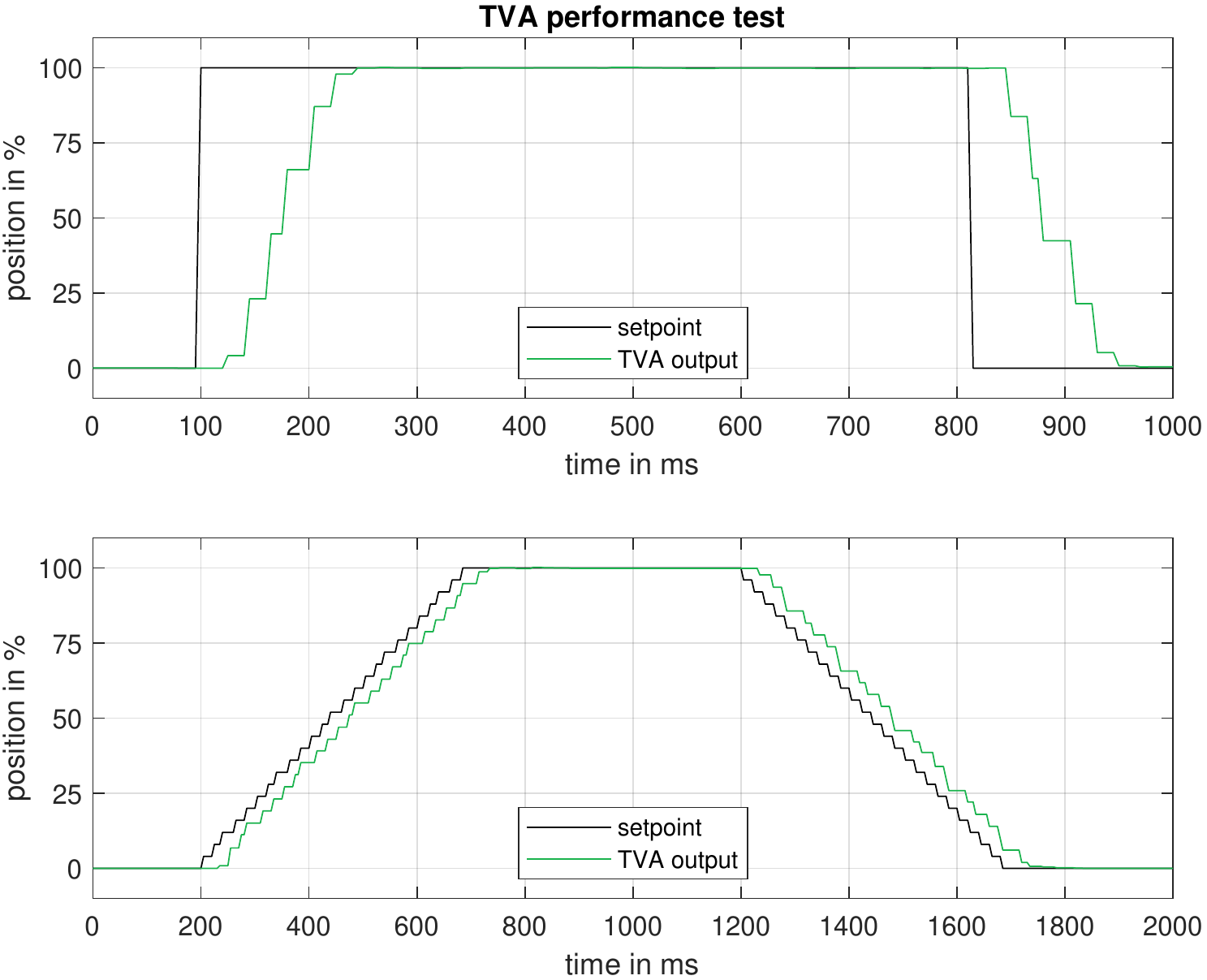}
\caption{\ac{TVA} performance evaluation}
\label{TVA_eval}
\end{figure}

The \ac{TVA} can be evaluated by using its already build in closed-loop control while being installed at the throttle valve. For realistic dead time behavior, the set valve positions were sent via rest bus simulation. In order to make qualitative statements about the control performance, the \ac{TVA} was excited with a pulse function and a sequence of two ramps. The dynamics of a step response cannot be generated by the rider, but represent a standard approach to control performance evaluation. Figure \ref{TVA_eval} shows the recorded scenarios. The \ac{TVA} is able to adjust within 130ms at max. control deviation, including bus communication. An overshoot of less than 0.5\% occurs and a stationary deviation of max. 0.1\%, excluding Hall sensor error of less than 0.37\%, is achieved. 
 

\subsection{Evaluation of the \ac{TbWS}}
Finally, the developed \ac{TPS} and \ac{TVA} had been integrated into the \textit{Peugeot Kisbee 50 4T (Euro 5)} scooter. In combination they can operate as \ac{TbWS}, open up the possibility of throttle manipulation through driving functions. The \ac{TPS} was placed beside the helmet compartment in such way, that the original throttle cable can be inserted. By mounting the \ac{TVA} directly at the engines air intake manifold and connecting both units with the power supply and the CAN bus, the \ac{TbWS} is ready to operate. 
\begin{figure*}[!b]
\centering
\includegraphics[width=14cm]{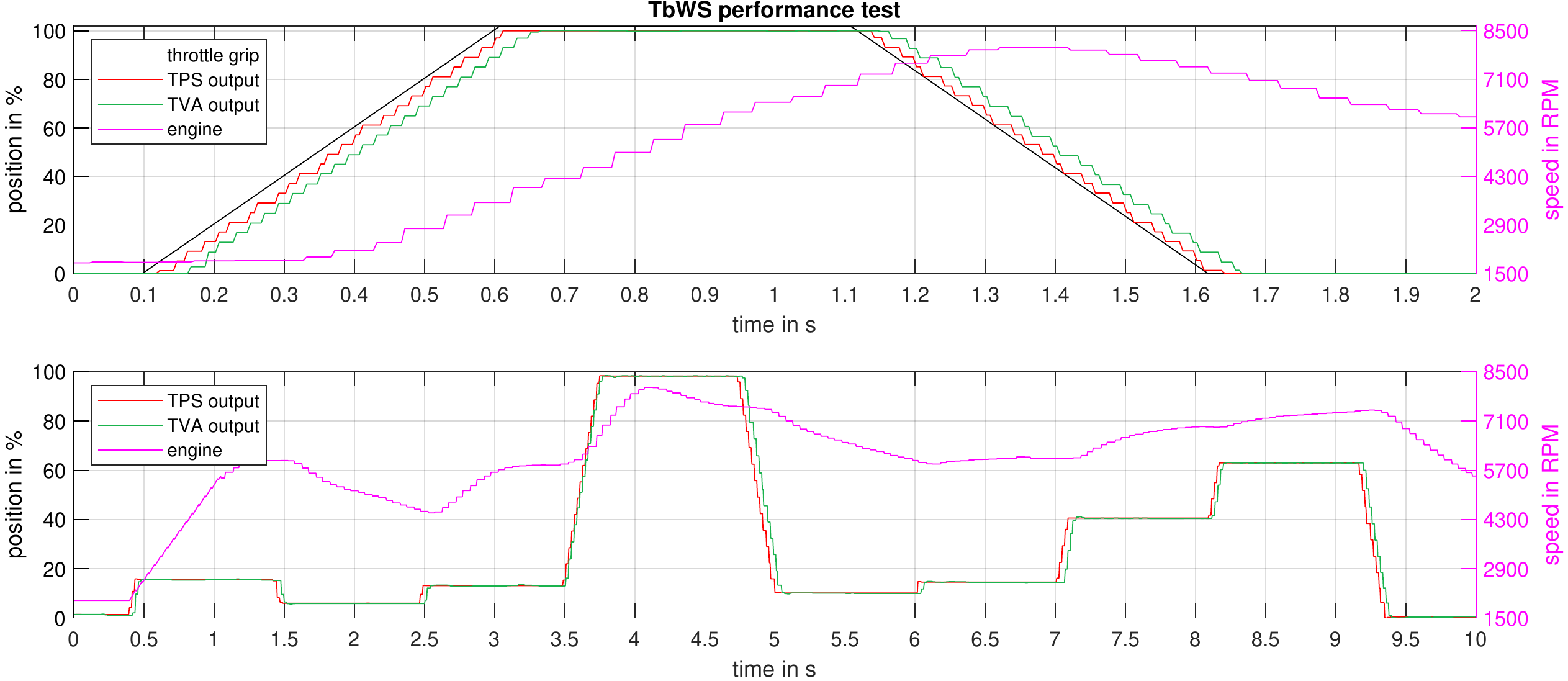}
\caption{\ac{TbWS} performance evaluation}
\label{TbWS_eval}
\end{figure*}
For testing the whole chain of moved cable pulley up to the actuated throttle valve/resulting engine behavior, the system was tested with a ramp sequence. During the test, the scooters engine was running for taking into account the suction pressure applied to the throttle valve. This time, the stepper motor is directly turning the \ac{TPS} cable pulley by means of an ideal straight-line throttle actuation. Figure \ref{TbWS_eval} shows the belonging results of applied throttle grip position, \ac{TPS}-measured throttle grip position, \ac{TVA}-actuated throttle valve position and additionally the engine speed. The first ramp-based test run performed, proves an excellent measurement of the throttle position, proper CAN communication and precise position control. The ramp used here, corresponds approximately to a rapid full throttle by the driver and subsequent release. After less than 30ms, the valid result of the \ac{TPS} was sent via CAN bus. The real time offset is less, but due to the test setup an offset between motor control and CAN message reception occurs. When reaching the 100\% position, a deviation of 1\% can be seen by considering the 1\% \ac{TPS}-error band. The recorded values reflect the CAN refreshing rate of 50Hz. The \ac{TVA} also follows the required position value within 30ms, thus the overall dead time is about less than 60ms. The position accuracy achieved is max. 0.1\%. It can be seen that even when higher engine speeds/various suction pressures are reached, exact position control by the \ac{TVA} is proven. For evaluation under extreme dynamics, open-loop and rapidly successive random throttle grip positions were controlled in a second test. The rear brake was lightly pressed to apply a uniform load to the powertrain. The engine speed exploited the full spectrum of range and dynamics. It can be seen that the accuracy of the \ac{TVA} is minimally dependent on engine speed. For low and medium speeds, an accuracy of min. 99.9\% is achieved. At higher speeds, strong vibrations lead to an accuracy of min. 99.8\%. Under consideration of the Hall sensor error ($<$0.27\%) and \ac{TPS} accuracy ($<$99.84\%), an overall error of max. 0.63\% results. The adjustment is always achieved within 70ms and the \ac{TVA} is capable of withstanding the torque variations caused by the changing suction pressure.\\

Besides performance, the system also meets low-cost requirements. Component and manufacturing costs for the prototypes of \ac{TPS} and \ac{TVA} amount to approx. 45€ and 95€, respectively. Also, instead of development boards, the MCU would be soldered directly to the PCB. In combination with further miniaturization, a single-unit design conceivable, eliminating the need for one MCU and reducing costs by about 25\%. Experience shows that costs in mass production are reduced by up to 90\%, due to larger purchase quantities and optimized processes. Consequently, final estimated total costs of min. 15€ per system could be realistic.

\section{Discussion}
In order to relate the \ac{TbWS} results to the context of other research, table \ref{Tab_Compare} gives a performance comparison to several electronic throttles (ETs) . Therefore, position error, settling time (ST) and overshoot behavior are investigated. All three reference systems are based on a PID-controlled friction compensated DC motor for use in motorcycles. The evaluation of referred systems included only the \ac{TVA}, which is why the direct comparison is made with the \ac{TVA} presented here. Within a benchmark, the developed \ac{TVA} performs best in both accuracy and overshoot performance. In terms of settling time, the system matches the performance of its counterparts. These were designed for use on racing motorcycles, while the requirement on a scooter is much lower. Considering the resulting system performance and the less complex design, the \ac{TbWS} performs very well. The maximum \ac{TbWS} accuracy includes the non-linearity of both position measurements, in contrast to the specifications of the compared systems.\newpage

\begin{table}[h]
\begin{center}
\caption{Performance comparison}
\label{Tab_Compare}
\begin{tabular}{l c c c c c }
\hline & \ac{TVA}  & ET1\cite{TVA_TbW_Ctr_SportMotorbike} & ET2 \cite{TVA_ETC_SportMotorbike} & ET3 \cite{TVA_ServoCtrDesign} &  \textbf{\ac{TbWS}} \\ \hline
Controller     &  PI     & PID    & PID      & PID    &  \textbf{PI}     \\
Motor          &  Step.  & DC     & DC       & DC     &  \textbf{Step.}  \\ 
Error (\%)     &  0.1    & 0.15   & 0.15     & 0.69   & \textbf{0.63}    \\ 
ST, step (ms)  &  130    & 120    & -        & 250$^*$&  \textbf{130}    \\ 
ST, ramp (ms)  &  30     & -      & 20       & -      &  \textbf{60}     \\
Overshoot (\%) &  \textless{}0.5  & \textless{}1.0 & \textless{}0.5  & \textless{}1.0  &  \textbf{\textless{}0.5} \\ \hline
\multicolumn{6}{r}{$^*$step size = 10\% of the opening range }
\end{tabular}
\end{center}
\end{table}

\vspace{-0.1cm}
By merging both units, the integration effort could be reduced and the system architecture simplified. Assemblies would have to be significantly reduced in size and the use of a smaller stepper motor would have to be considered. The decisive factor would be the development of a central PCB and the elimination of one \ac{MCU}. The temperature dependence of the entire system could be evaluated in a climate chamber with regard to extreme temperatures.

\section{Conclusion}
In this paper, the development of a low-cost \ac{TbWS} is treated, consisting of a sensor unit determining the throttle grip position and an actuator manipulating the throttle valve position. The throttle grip position was measured redundantly and contactless by means of an \ac{AMR} sensor. With the help of plausibility algorithms, the proper function of the \ac{TPS} can be verified. For easy vehicle integration/retrofit, the existing throttle cable is directly attached to the unit. The digital measured value is shared via CAN bus with the \ac{TVA}, which sets the required position at the throttle valve. The actuator consists of a stepper motor-based drivetrain and the position is controlled closed-loop using position measurement as feedback. By using a high performance motor driver, very accurate positions can be approached while vibrations and acoustic noise are kept to a minimum. A PI-controller is able to control precisely without taking friction compensation into account or working model-based. When an error occurs, the subsystems are able to differentiate between different faults. Problems with CAN communication, position measurement or faulty throttle positions can be detected. Both units are able to request calibration or perform them independently. In case of major errors, the ignition can be interrupted to guarantee safe condition. The accuracy of the \ac{TbWS} is 99.37\% of the throttle grip position and is reached within approx. 60ms, which corresponds to a much faster reaction than the response of a small-volume single-cylinder four-stroke engine. For the prototype production costs of about 140€ have been incurred. These can be significantly reduced through improvements in size and mass production, lowering the cost of manufacturing.\\

A low-cost \ac{TbWS} allows effective usage in scooters and small motorcycles, enabling implementation of ADAS. Scooters in particular are used by millions and could be restricted more ecologically and contribute to environmental protection. Effects on fuel consumption, associated CO$_{2}$ emissions and improved exhaust gas composition are to be expected. A demonstration video was created, which can be accessed at the link: \url{https://youtu.be/6Fzch4g0zOE}


 %


\vspace{11pt}

\vspace{-33pt}
\begin{IEEEbiographynophoto}{Jannis Kreß} received the B.Eng. degree in mechatronics in 2018 and M.Sc. degree in mechatronics \& robotics in 2019 from Frankfurt University of Applied Sciences, Germany. He is currently working toward the Ph.D. degree regarding intelligent throttling of two-wheeler by using Drive-by-Wire-Systems at University of Cadiz, Spain.  
\end{IEEEbiographynophoto}
\vspace{-33pt}
\begin{IEEEbiographynophoto}{Jens Rau} received the B.Eng. degree in mechatronics in 2022 from Frankfurt University of Applied Sciences, Germany. He is currently working towards the M.Sc. degree in mechatronics \& automotive engineering besides being a student research assistant at Frankfurt University of Applied Sciences. 
\end{IEEEbiographynophoto}
\vspace{-33pt}
\begin{IEEEbiographynophoto}{Hektor Hebert} Hektor Hebert received the diploma in physics from Goethe University in Frankfurt and the Ph.D. from the Humboldt University in Berlin. From 1994 to 2013, he worked with several research and development organizations in Europe, Asia and the USA. Since 2013 he is a professor for physics and mechatronics at Frankfurt University of Applied Sciences.
\end{IEEEbiographynophoto}
\vspace{-33pt}
\begin{IEEEbiographynophoto}{Fernando Perez-Peña} received the Engineering degree in Telecommunications from the University of Seville (Spain) and his Ph.D. degree (specialized in neuromorphic motor control) from the University of Cadiz (Spain) in 2009 and 2014 respectively. In 2015 he was a postdoc at CITEC (Bielefeld University, Germany). He has been an Assistant Professor in the Architecture and Technology of Computers Department of the University of Cadiz since 2014. His research interests include neuromorphic engineering, CPG, motor control and neurorobotics.
\end{IEEEbiographynophoto}
\vspace{-33pt}
\begin{IEEEbiographynophoto}{Karsten Schmidt} received his Ph.D. in nuclear physics in 1995 and gained experience in IT and automotive companies. Since 2005, he has been a professor for mechatronics, first at the Coburg University of Applied Sciences and later at the Frankfurt University of Applied Sciences. His research interests are embedded solutions for robotics and automotive applications.
\end{IEEEbiographynophoto}
\vspace{-33pt}
\begin{IEEEbiographynophoto}{Arturo Morgado Estévez} received his PhD in Industrial Engineering from the University of Cadiz (UCA, Spain) in 2003. He started as an interim professor in 1987 and has been an assistant professor at UCA from 1991 to the present. He was deputy head of research/innovation and transfer of the faculty of engineering (2004-2014) and is head of the TEP-940 Applied Robotics Research Group. His current research interests are industrial and educational robotics, robotics hardware/software prototyping and electronics design for embedded systems.
\end{IEEEbiographynophoto}

\vfill

\end{document}